\documentclass[aps,physrev,preprint,groupedaddress]{revtex4-2}

\usepackage{newtxtext,newtxmath}
\usepackage{babel}
\usepackage{CJKfntef,CJK}

\usepackage[T1]{fontenc}
\usepackage{verbatim}
\usepackage{color}
\usepackage{url}
\usepackage{subcaption}
\usepackage{natbib}
\usepackage{hyperref}
\usepackage{graphicx}	
\usepackage{amsmath}	
\usepackage{siunitx}
\usepackage{threeparttable}
\usepackage{ulem}
\usepackage{array}
\usepackage{multirow}

\DeclareRobustCommand{\VAN}[3]{#2}
\let\VANthebibliography\thebibliography
\def\thebibliography{\DeclareRobustCommand{\VAN}[3]{##3}\VANthebibliography}

\newcommand{\s}{\,{\rm s}}      
    
\newcommand{\cm}{\,{\rm cm}}    \newcommand{\km}{\,{\rm km}}

\newcommand{\erg}{\,{\rm erg}}

\newcommand{\gray}{$\gamma$-ray}

\begin{document}
\begin{CJK*}{UTF8}{bsmi}

\title{On the MeV Nuclear De-excitation Line Emission from the Vela Supernova Remnant Region}

\author{Yun-Zhi Shen (沈蘊之)}
\affiliation{School of Astronomy \& Space Science, Nanjing University, 163 Xianlin Avenue, Nanjing 210023, China}

\author{Bing Liu (劉冰)}
\email[]{liubing@pmo.ac.cn}
\thanks{}
\affiliation{Purple Mountain Observatory, Chinese Academy of Sciences, Nanjing 210023,  China}

\author{Yang Chen (陳陽)}
\email[]{ygchen@nju.edu.cn}
\thanks{}
\affiliation{School of Astronomy \& Space Science, Nanjing University, 163 Xianlin Avenue, Nanjing 210023, China}
\affiliation{Key Laboratory of Modern Astronomy and Astrophysics, Nanjing University, Ministry of Education, Nanjing 210023, China}


\date{\today}

\begin{abstract}
We investigate the potential nuclear de-excitation $\gamma$-ray line emission from low-energy cosmic rays (LECRs) in the Vela SNR region, one of the nearest and most angularly extended core-collapse supernova remnant (SNR) complexes.
We calculate the MeV line emission from Vela, Vela Jr., and Puppis A using the TALYS nuclear reaction code, considering both direct and inverse interactions between the accelerated particles and the nuclei interior to the SNRs.
The proton spectra are constrained by GeV--TeV $\gamma$-ray observations under the hadronic assumption, while a broad range of spectral indices is explored to account for the poorly constrained LECR population.
We find that Vela dominates the predicted line flux among the three SNRs.
The integrated fluxes of the prominent 1.63 MeV $^{20}$Ne, 4.44 MeV $^{12}$C and 6.13 MeV $^{16}$O lines are approximately $1.3\times10^{-10}$ -- $1.2\times10^{-6}$, $3.2\times10^{-10}$ -- $5.6\times10^{-7}$ and $1.6\times10^{-10}$--$2.6\times10^{-7}$ $\rm ph\,cm^{-2}\,s^{-1}$, respectively, over the explored proton spectra.
We compare the predicted line emission with both the diffuse Galactic MeV continuum and local source-related continuum emission.
In our background estimate, diffuse Galactic emission is the dominant component, while inverse Compton emission from pulsar wind nebula Vela X constitutes an additional local continuum component.
The relatively low diffuse background toward Vela compared with the Galactic Center makes this region a promising target for future MeV $\gamma$-ray observations.
Such observations could provide direct constraints on the LECR population and particle acceleration in this complex SNR region.
\end{abstract}


\maketitle

\section{Introduction}
Cosmic rays (CRs) interact with the interstellar medium (ISM) through a variety of energy-dependent processes, producing distinct signatures across the electromagnetic spectrum.
At energies above the pion-production threshold ($\sim 280$ MeV), high-energy protons can interact with ambient gas and produce high-energy $\gamma$-rays through hadronic processes.
In contrast, low-energy CRs (LECRs) below this threshold primarily contribute to ionization of the ISM, thereby influencing interstellar chemistry and thermal balance \citep{CR280MeV, 280MeV_chem}. 
In addition, LECRs can inelastically interact
with the nuclei they encounter, exciting the nuclei to higher-energy states, which subsequently de-excite by emitting characteristic nuclear $\gamma$-ray lines spanning energies from tens of keV to $\sim20$ MeV, including the prominent 4.44 MeV line from $^{12}$C and the 6.13 MeV line from $^{16}$O \citep{RKL1979}.
These nuclear de-excitation lines provide a unique diagnostic of LECR populations that cannot be directly probed by in situ measurements, offering a direct probe of the low-energy end of the hadronic CR spectrum.
Nuclear $\gamma$-ray line emission has also been investigated in other astrophysical environments, such as solar flares \citep{Murphy2009}.
Direct measurements of LECRs in the heliosphere are significantly affected by solar modulation effects, although the Voyager mission has provided constraints on the interstellar CR spectrum beyond the heliopause \citep{voyager}.
However, Galactic LECRs are expected to be distributed highly non-uniformly across the Galactic plane owing to their rapid energy losses and limited propagation distances, while their spatial distribution remains poorly constrained observationally \citep{Indriolo2015}.
Consequently, regions with enhanced CR ionization rates, such as the central molecular zone \citep{cmz_nline}, are particularly promising targets for searches for LECR-induced nuclear de-excitation lines.

Supernova remnants (SNRs) are widely regarded as the primary accelerators of Galactic CRs. GeV–TeV observations have demonstrated that SNRs can accelerate particles up to very high energies, supporting their role as efficient proton accelerators \citep[e.g., IC443, and W44,][]{IC443_W44_gray}. 
In addition, enhanced ionization rates typically two to three orders of magnitude above the Galactic average, have been observed in molecular clouds interacting with SNRs, providing indirect evidence for substantial populations of LECRs associated with SNRs, 
\citep[e.g., IC 443, W28, and W49B,][]{ic443, W28_V, W28_Tu, w49b}.
Thus, the observations of the MeV \gray\ lines will derive the unique information about the injection \citep[e.g., ][]{de_excitation_line_liub, de_excitation_casa}.

Vela (G263.9$-$3.3) is one of the nearest core-collapse SNRs, at a distance of approximately 290 pc \citep{Vela_dis}.
It spans a large angular extent with a radius of about 4.1$^\circ$ \citep{Vela_radio} and contains the pulsar wind nebula (PWN) Vela X powered by PSR B0833$-$45.
In addition, the same region of the sky also includes two other SNRs, Vela Junior (G266.2$-$1.2; hereafter Vela Jr.) and Puppis A (G260.4$-$3.4), located at distances of approximately 750 pc and 1.3 kpc, with smaller angular sizes of about 1$^\circ$ and 0.5$^\circ$, respectively.
The elemental abundances of Vela have also been relatively well constrained by X-ray observations \citep{ele_abundance_Vela}, providing useful constraints on the target composition for nuclear de-excitation line production. 
These properties make the Vela SNR region a particularly suitable laboratory for studying LECR-induced MeV $\gamma$-ray lines.
At high energies, these sources exhibit diverse emission properties. The Vela X PWN shows a multi-component leptonic spectrum extending from GeV to TeV energies \citep[e.g., ][]{VelaX_fermi_2025, VelaX_Bao1, VelaX_Bao2}.
The broader Vela shell emission observed in the GeV band indicates a large extended component likely associated with hadronic CRs \citep{Vela_GeV}.
Vela Jr. shows evidence for both leptonic and hadronic scenarios, while the hard spectrum of power-law index $\alpha \sim 1.8$ in the hadronic case suggests efficient particle acceleration \citep{VelaJr_gray}.
GeV-TeV $\gamma$-rays from Puppis A are also consistent with hadronic emission possibly related to escaped CRs interacting with surrounding molecular clouds \citep{puppisa_GeV}.
These observations indicate that the Vela SNR region hosts multiple populations of accelerated particles.
Vela SNR region provides a nearby, spatially extended system with multiple overlapping accelerators.
Moreover, its large angular extent is comparable to the angular resolution of next-generation MeV telescopes, making it a promising target for spatially resolved studies of LECRs \citep{newASTROGAM2025, COSI2023}.
The coexistence of the Vela shell, Vela X, Vela Jr., and Puppis A further offers a unique opportunity to investigate both hadronic and leptonic particle populations within a single MeV field of view. 
However, the low-energy component of the CR spectrum, which is crucial for nuclear de-excitation line production, remains largely unconstrained. 
MeV $\gamma$-ray observations therefore provide a unique probe of the low-energy particle population and may help disentangle the contributions from different acceleration sites in this complex region.

In this work, we first investigate the potential MeV nuclear de-excitation line emission from the Vela SNR region under a set of assumptions for the injected CR spectra, constrained by GeV--TeV $\gamma$-ray observations, as well as elemental abundances inferred from X-ray studies.
The adopted CR spectral models and interacting medium properties are described in Section~\ref{sec: CR spectrum}.
The resulting predictions for MeV de-excitation line emission are presented in Section~\ref{sec: results}.
In Section~\ref{sec: discussion}, we further discuss the detectability of these lines in the presence of diffuse Galactic emission and continuum contributions from the SNR environment itself.

\section{CR spectrum and the medium composition in Vela Region}\label{sec: CR spectrum}

\begin{figure}
    \includegraphics[width=0.8\textwidth]{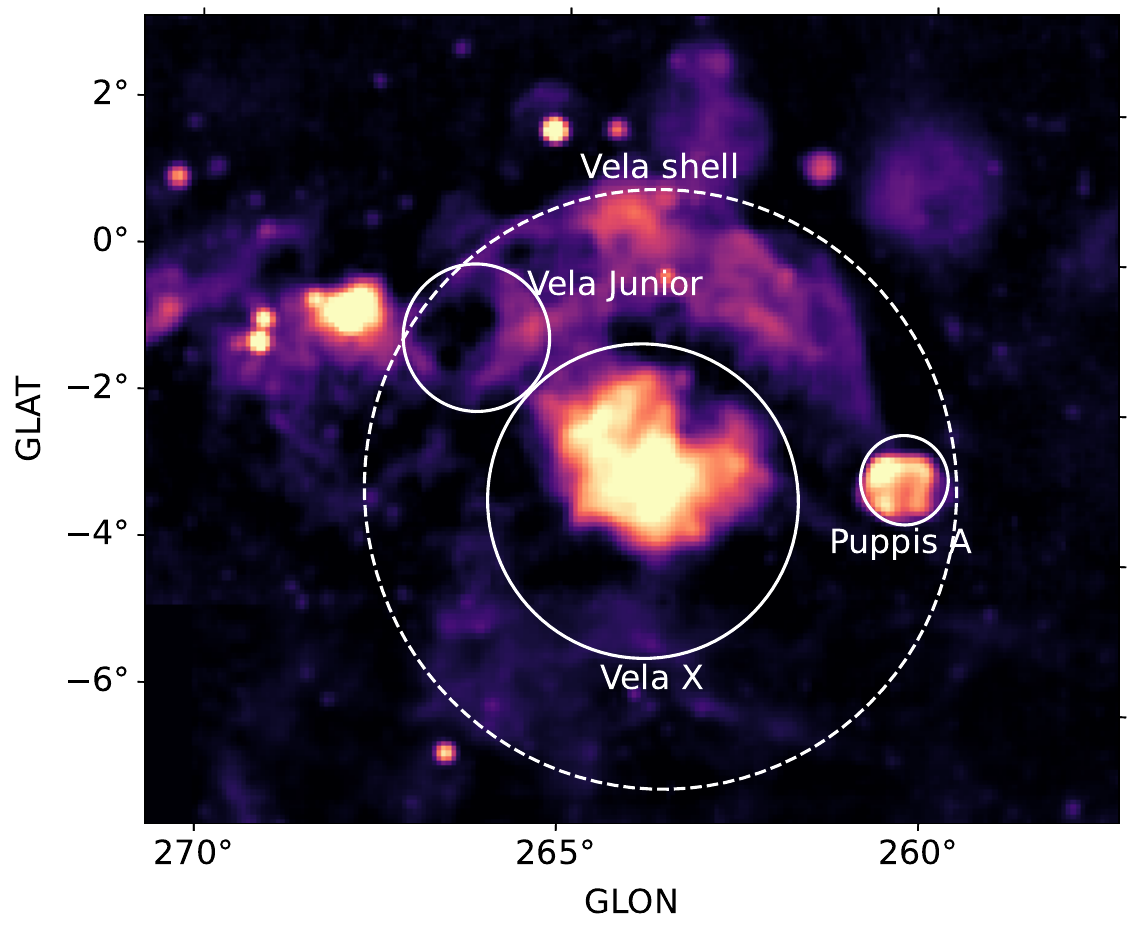}
    \caption{The radio emission in the region of SNR Vela, taken from \citet{Vela_radio}.
    }
    \label{fig: vela_radio}
\end{figure}

Figure~\ref{fig: vela_radio} shows the radio emission in the Vela SNR region. 
According to \citet{Vela_radio}, the boundary of the Vela shell is outlined by the dashed white circle with a radius of $\sim4.1^\circ$, mostly overlapping with the other two SNRs, Vela Junior and Puppis A.
Motivated by the limited imaging capability at MeV energies, as well as the substantial uncertainties in the spatial distribution of LECRs and their coupling to the target material within the remnant, we adopt a spatially integrated approximation for the remnant and estimate the total de-excitation line emission integrated over the entire SNR.
Similar assumptions are adopted for Vela Junior and Puppis A, whose smaller angular sizes and larger distances further limit the possibility of resolving internal structures in the MeV band.

\subsection{Spectral distribution of SNR accelerated LECR particles}

Adopting the energy-loss rate provided by Figure~7 of \citet{loss_rate} and the deduction from \citet{de_excitation_line_liub}, the energy-loss timescale is $\sim10^4$ yr for 1 MeV protons assuming the medium density $n=1~\cm^{-3}$ and $\sim10^5$ yr for 10 MeV protons, which dominate the production of the 4.44 and 6.13 MeV nuclear de-excitation lines of $^{12}$C and $^{16}$O, respectively. Thus, given the estimated age of Vela SNR  \citep[$\sim10^4$ yr; e.g.,][]{velaage1, velaage2, velaage3}, Vela Jr. \citep[2.4 -- 5.1 kyr,][]{Velajr_age} and Puppis A \citep[approximately 2.2 -- 4.4 kyr,][]{PuppisA_age}, the energy losses are neglected in our calculations for LECR components from Vela, Vela Jr., and Puppis A.
In the framework of diffusive shock acceleration, the resulting particle spectra are often approximated as single power laws in momentum or energy.
Since energy losses are not expected to significantly modify the proton spectrum in the energy range relevant for nuclear de-excitation line production, we model the injected proton population using a power law with an exponential cutoff:
\begin{equation}
\begin{aligned}
    dN/dE =A (E / 1\ {\rm GeV})^{-\alpha} \exp(-E/E_{{\rm cut}}),
\end{aligned}
\end{equation}
where $A$ , $\alpha$, $E_{{\rm cut}}$ and $E$ are the normalization factor, spectral index, the cutoff energy, and the proton kinetic energy, respectively.

For the three SNRs considered in this work, the proton spectral parameters are constrained by their observed GeV--TeV $\gamma$-ray spectra, with the adopted values summarized in Table~\ref{tab: spepars}.
The left panel of Figure 2 shows the spectral energy distribution (SED) of the Vela SNR measured by Araya et al. \citep{Vela_GeV}, together with illustrative pion-decay models.
The right panel shows the best-fit pion-decay SEDs of Vela Jr. and Puppis A.
We use these GeV -- TeV derived proton spectra as the basis for estimating the MeV line emission, while exploring a broader range of spectral indices to account for the uncertainty in extrapolating the proton spectra from the GeV -- TeV range to lower energies.
Motivated by the hadronic interpretation of Vela's GeV emission presented by \citet{Vela_GeV}, we adopt $\alpha=3.0$ as the reference index for our kinetic-energy proton spectrum and additionally consider $\alpha=4.0$ to explore a softer low-energy extrapolation.
For Vela Jr. and Puppis A, we adopt the value of $\alpha=1.8$ from their respective $\gamma$-ray studies and additionally explore $\alpha=2.0$, 2.3, and 2.6. 
The remaining spectral parameters, including the cutoff energy $E_{\rm cut}$, are listed in Table~\ref{tab: spepars}.
The proton spectral normalization is constrained by the observed GeV -- TeV $\gamma$-ray emission under the assumption of a hadronic origin, namely pion-decay emission produced by interactions between accelerated protons and ambient gas.
The $\gamma$-ray spectral data are adopted from \citet{Vela_GeV, VelaJr_gray, puppisa_GeV}.
Vela Jr. likely has a substantial leptonic contribution to its GeV--TeV emission \citep{VelaJr_gray}.
For the adopted proton spectra, the fully hadronic interpretation used here therefore represents an optimistic scenario for its nuclear de-excitation line emission.
A smaller hadronic contribution would reduce the predicted line fluxes.
To examine the uncertainty in extrapolating the GeV-derived proton spectrum to lower energies, we additionally consider a broken power law (BPL) with a harder low-energy segment for Vela.
We adopt the following exponentially cut-off broken power law for the proton spectrum:
\begin{equation}
\frac{dN}{dE}
= A_{\rm BPL}\exp\!\left(-E/E_{\rm cut}\right)
\begin{cases}
(E/E_0)^{-\alpha_1}, & E<E_{\rm b},\\[2pt]
(E_{\rm b}/E_0)^{\alpha_2-\alpha_1}
(E/E_0)^{-\alpha_2}, & E\geq E_{\rm b},
\end{cases}
\label{eq:vela_bpl}
\end{equation}
where $E$ denotes proton kinetic energy and $E_0=1$~GeV.
Fixing $\alpha_1=2.2$ and restricting $E_{\rm b}$ to 5--10~GeV, a fit to the published Fermi-LAT SED points yields $\alpha_2\simeq2.74$, with $E_{\rm b}=10$~GeV at the upper boundary.
For the MeV predictions, we renormalize the fitted shape to $W_p=4\times10^{49}$~erg over 0.01--$10^5$~GeV, retaining $E_{\rm cut}=10^6$~GeV and the other Vela parameters in Table~I.

\begin{table}
    \centering
    \caption{Spectra and environmental parameters for the three SNRs.}
    \label{tab: spepars}
    \begin{tabular}{ccccccc}
    \hline
    \hline
          SNR  &     $E_{\rm cut}^a$ & Proton Energy  &  Distance$^b$ & Radius  & Total mass$^c$\\
               &   (GeV) & ($10^{49} \erg$) &  (pc)   &  (pc)  & ($M_{\odot}$) \\   
    \hline
        Vela  &  $1\times10^6$  & 4 & 290  & 21  &  200 \\
        Vela Junior & $5.5\times10^4$  & 7 & 750  &  11  & 40 \\
        Puppis A  &  $1.0 \times 10^3$   & 1 & 1300  &  11  &  80\\
    \hline
    \end{tabular}
    \begin{tablenotes}
        \small
        \item $^a$ Cutoff energies $E_{\rm cut}$ adopted from \citet{Vela_GeV, VelaJr_gray, puppisa_GeV}.
        \item $^b$ Distances adopted from \citet{Vela_dis, velajr_dis, puppisa_dis}.
        \item $^c$ Gas masses inside the SNRs estimated using the X-ray analyses of \citet{ele_abundance_Vela, ele_abundance_VelaJr, abun_puppisa}.
        
    \end{tablenotes}
\end{table}

\begin{figure*}
    \includegraphics[width=0.95\textwidth]{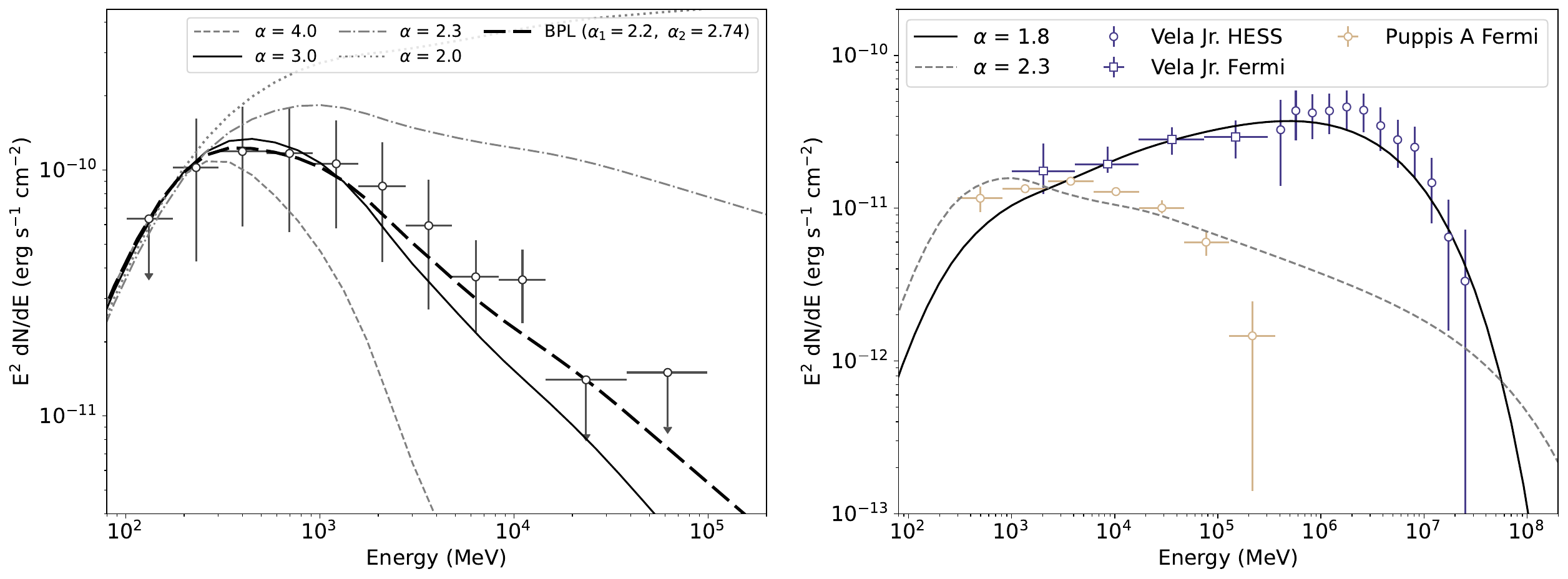}
    \caption{
    \textit{Left:} Fermi-LAT SED of the Vela SNR reported by Araya et al. \cite{Vela_GeV}, together with illustrative pion-decay models for different assumed proton spectra. The BPL model has kinetic-energy indices $\alpha_1=2.2$ and $\alpha_2=2.74$, with a break at 10 GeV.}
    \textit{Right:} SEDs of the Vela Jr. and Puppis A SNRs with the best-fit pion-decay model taken from \citet{VelaJr_gray} and \citet{puppisa_GeV}.
    
    \label{fig: sed_vela}
\end{figure*}

\subsection{Elemental abundances of the SNR material}

For the three SNRs considered in this work, we treat their elemental abundances individually using spatially resolved X-ray measurements \citep{ele_abundance_Vela, ele_abundance_VelaJr, abun_puppisa} and adopt the solar abundance value for carbon of Vela and Puppis A. 
Since the reported abundances vary across different regions of each remnant, we estimate a remnant-averaged abundance weighted by the hydrogen content of the X-ray emitting gas:
\begin{equation}
(X/X_\odot)_{\rm total} = \frac{\Sigma_i(X/X_\odot)_i N_{{\rm H}, i}}{\Sigma_i N_{{\rm H}, i}},
\end{equation}
where $X$ is the elemental abundance and $N_{{\rm H}, i}$ is the hydrogen in each region $i$. 
Here, $N_{{\rm H}, i}$ is approximated as $\sqrt{EM_i V_i}$, in which, $V_i$ and $EM_i$ are the volume and the volume emission measure for region $i$,  respectively.

Due to the lack of reliable geometric information for individual regions, we assume comparable emitting volumes among regions and use $\sqrt{EM_i}$ as an approximate weighting factor \footnote{Adopting equal volumes mainly affects the absolute abundance normalization but is unlikely to qualitatively alter the relative elemental abundance trends.}.
Since the elemental composition of accelerated LECRs in SNRs is poorly constrained observationally, we adopt the Voyager measurements of local LECR abundances as a reference assumption for the accelerated particle composition in all three remnants \citep{voyager}. 
All averaged elemental abundance ratios adopted in this work are summarized in Table~\ref{tab: abundance}.

\begin{table}
    \centering
    \caption{Elemental abundance ratios adopted in this work.}
    \label{tab: abundance}
    \begin{tabular}{cccccc}
    \hline
    \hline
            &   CR  &   Solar    &  Vela & Vela Jr.& Puppis A\\
            & $n_{\rm X}/n_{\rm H}$  &   $n_{\rm X} / n_{\rm H}$ &     $n_{\rm X} / n_{\odot}$ & $n_{\rm X} / n_{\odot}$  &  $n_{\rm X} / n_{\odot}$\\
    \hline
        H   &   1   &   1   &   1   &   1  & 1 \\
        He  & $8.140\times10^{-2}$ & $8.140\times10^{-2}$ & 1   & 1    & 1\\
        C   & $1.671\times10^{-3}$ & $2.455\times10^{-4}$ & 1   & 0.2  & 1\\
        N   & $2.444\times10^{-4}$ & $7.244\times10^{-5}$ & 0.4 & 0.1  & 0.4\\
        O   & $1.570\times10^{-3}$ & $5.370\times10^{-4}$ & 1.1 & 1.4  & 2.5\\
        Ne  & $1.507\times10^{-4}$ & $1.122\times10^{-4}$ & 2.2 & 0.8  & 2.7\\
        Mg  & $2.264\times10^{-4}$ & $3.467\times10^{-5}$ & 2.0 & 0.6  & 2.8\\
        Si  & $1.898\times10^{-4}$ & $3.388\times10^{-5}$ & 3.6 & 0.3  & 4.5\\
        S   & $2.087\times10^{-5}$ & $1.445\times10^{-5}$ & 0   & 0    & 4.8\\
        Fe  & $1.152\times10^{-4}$ & $2.884\times10^{-5}$ & 1.5 & 0.1  & 0.7\\
    \hline
    \end{tabular}
\end{table}

We use the electron density to estimate the mass of the X-ray emitting plasma. 
For example, \citet{ele_abundance_Vela} derive an electron density of $n_{\rm e}\sim2~{\rm cm^{-3}}$ for the Vela SNR, corresponding to a hydrogen density of $n_{\rm H}\sim n_{\rm e}$, while \citet{Vela_dens2} adopt a density of $0.5~\cm^{-3}$. 
We therefore adopt a representative value of $n_{\rm H} = 1~{\rm cm^{-3}}$ for calculations of Vela SNR.
For SNR Vela Jr., we similarly adopt $n_{\rm H} = 1 \cm^{-3}$ as an estimate.
Assuming a fully ionized plasma with approximately solar composition, the total mass  of the emitting material $M$ is estimated as $m_{\rm p} n_{\rm e} 4\pi R^2 \Delta R$, where $m_{\rm p}$ is the proton mass, and $\Delta R \sim (1/12) R$.
Using this approach, we obtain the gas masses inside the SNRs of order $\sim 200$ M$_\odot$ for Vela and $\sim 40$ M$_\odot$ for Vela Jr., while for Puppis A we adopt the literature value of $\sim 80$ M$_\odot$ \citep{abun_puppisa}. 
These values are used as inputs for the calculation of MeV nuclear line emission.

The basic parameter framework has thus been constructed for the three targets considered in this work. 
A self-consistent description of the accelerated CR composition in such a complex region would require a comprehensive treatment of particle acceleration, escape, diffusion, and mixing processes, which is beyond the scope of the present study. 
Therefore, our calculations are intended to provide an order-of-magnitude estimate of the MeV de-excitation line fluxes in the Vela SNR region. 
More detailed modeling will become possible with next-generation MeV instruments capable of spatially resolved observations of these remnants.

\section{Possible de-excitation line emission in Vela SNR region}\label{sec: results}

\begin{figure*}
    \centering
    \includegraphics[width=1.0\textwidth]{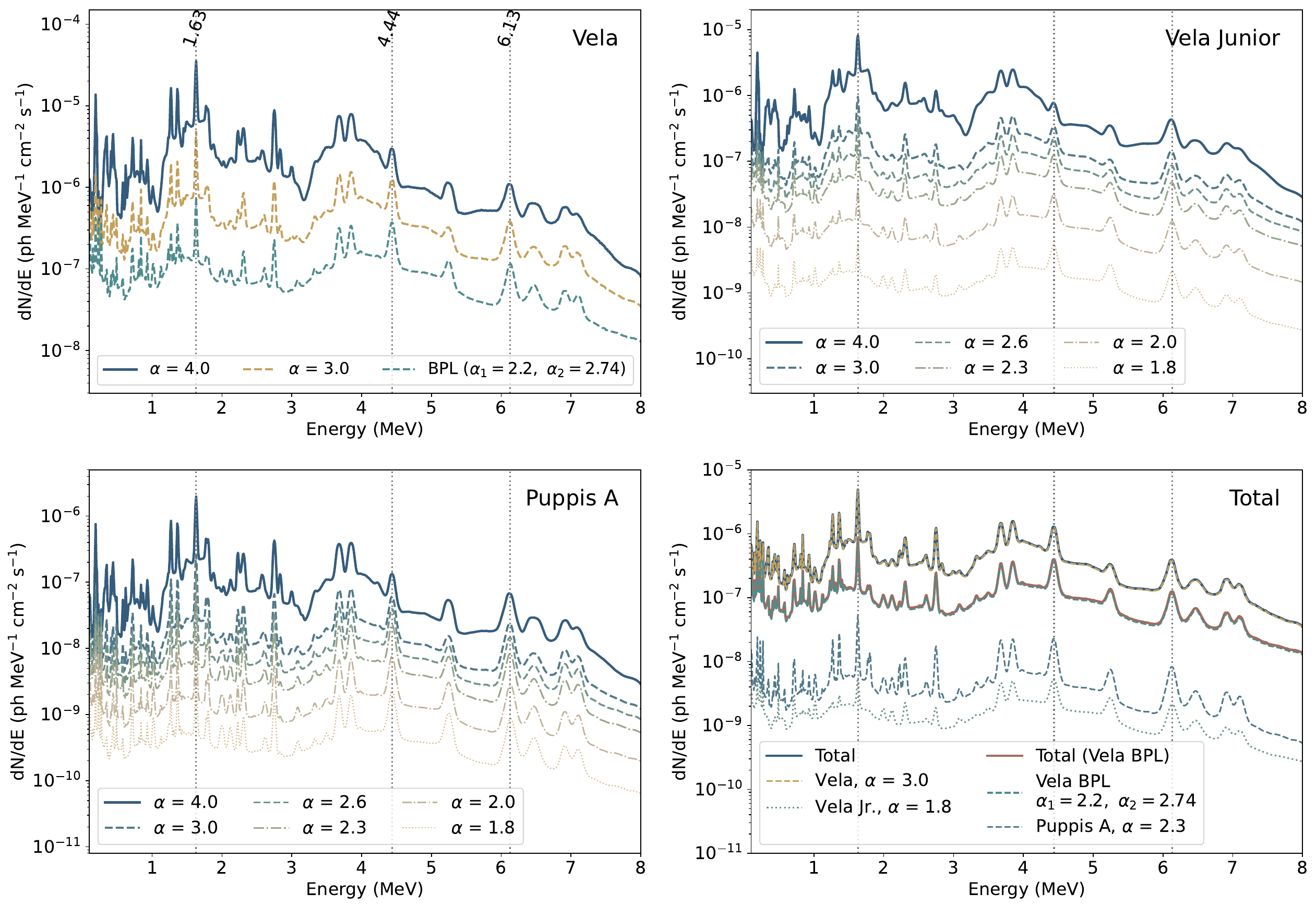}
    \caption{MeV nuclear de-excitation line differential spectra of the three target SNRs. The first, second, and third panels show the spectra of Vela, Vela Jr., and Puppis A, respectively, for different proton spectral indices $\alpha$. 
    The ``Vela'' panel also includes a BPL model with $\alpha_1=2.2$, $\alpha_2=2.74$, and a break at a proton kinetic energy of 10~GeV.
    The last panel show the total spectral emission of the three targets in the Vela SNR region, adopting $\alpha = 3.0$ for Vela, $\alpha = 1.8$ for Vela Jr., and $\alpha = 2.3$ for Puppis A, following the GeV -- TeV studies \citet{Vela_GeV, VelaJr_gray, puppisa_GeV}. 
    The alternative total spectrum replaces the Vela component with the BPL model. 
    In this panel, solid curves show the total spectra, while dashed and dotted curves show the individual SNR contributions.
    }
    \label{fig: all_spe}
\end{figure*}

We use version 2.2 of the nuclear reaction code {\textbf TALYS} to calculate the cross sections of proton- and $\alpha$-induced nuclear interactions relevant to MeV de-excitation line production \citep{talys}.
Since {\textbf TALYS} may generate de-excitation transitions associated with theoretically predicted or weakly constrained nuclear levels, we further cross-check the resulting line list using the Evaluated Nuclear Structure Data File (ENSDF), an experimentally based nuclear structure database, and exclude pseudo lines that are not supported by experimental measurements\footnote{\url{https://www.nndc.bnl.gov/ensdfarchivals/}}.
The line emissivity is calculated following the formalism described in Section 3.1 in \citet{de_excitation_line_liub}, which is based on the earlier developments of \citet{RKL1979, Murphy2009, BKT2013}.
We consider the two main interaction channels expected in the SNRs.
The first is the {\textit direct} process, in which accelerated protons and $\alpha$-particles in LECRs interact with heavy nuclei in the gas, exciting nuclear states and producing relatively narrow de-excitation lines.
The second is the {\textit inverse} process, in which the swept-up hydrogen and helium nuclei from the surrounding medium interact with accelerated heavy ions in LECRs, leading to broader lines due to the recoil motion of the excited nuclei.
The detailed line profiles are calculated following Equations (2)--(9) of \citet{RKL1979}, which include the effect of nuclear recoil kinematics and recoil angles on the emitted photon energies.

To derive the final MeV line profiles, we also include Doppler broadening associated with the bulk motion of the shocked ejecta. For a strong shock, the downstream velocity is approximated as $v_{\rm down} = (3/4) v_{\rm s}$, where $v_{\rm s}$ is the shock velocity. 
Assuming isotropic downstream motion, we estimate the corresponding Doppler broadening as $(E/c) (v_{\rm down} / \sqrt{3})$.
As an example, adopting $v_{\rm s}\sim 3000 \km\s^{-1}$ for Vela Jr. \citep{VelaJr_gray}, the resulting line broadening is approximately 30 keV, significantly smaller than the intrinsic line width induced by nuclear recoil. 
Therefore, Doppler broadening associated with the SNR shock dynamics is expected to have only a minor impact on the predicted MeV line profiles.

\begin{figure}
     \centering
     \includegraphics[width=0.8\textwidth]{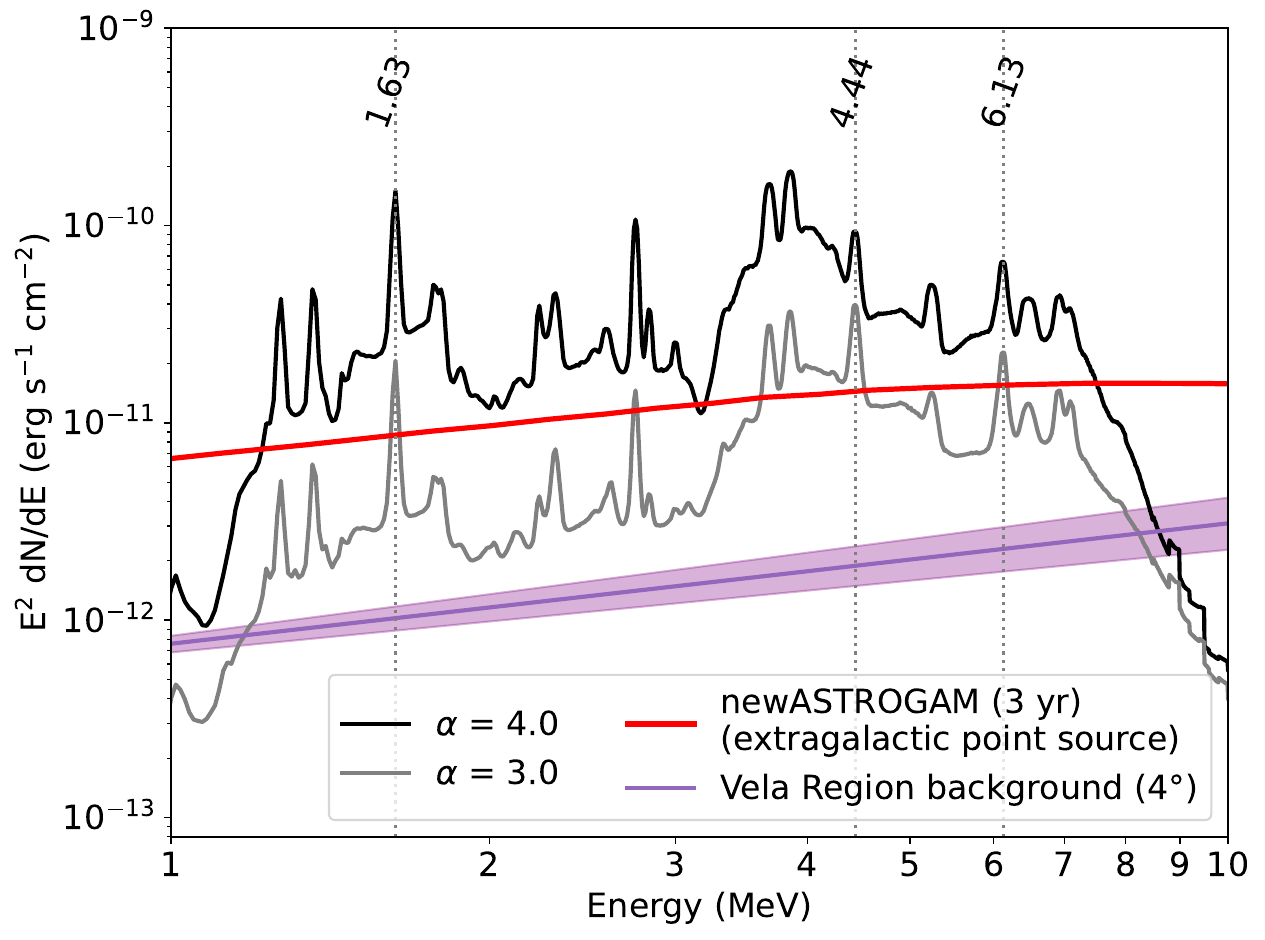}
    \caption{MeV nuclear de-excitation line spectra of the Vela SNR for different proton spectral indices, with $\alpha = 4.0$ (black), and $\alpha = 3.0$ (dark gray).
    The purple curve shows the diffuse continuum background estimated from the power-law model described in Section IV.A and integrated over a $4^\circ$-radius region, with the shaded band indicating its uncertainty.
    For reference, the continuum sensitivity of newASTROGAM for extragalactic point sources in a three-year survey is shown by the red curve, taken from Fig.~1 of \citet{newASTROGAM2025}.
    The black and gray curves show the intrinsic Vela nuclear emission spectra before instrumental convolution.
    }
    \label{fig: bkg}
\end{figure}

The superposition of contributions from ions with different energies and species leads to significant Doppler effect of the recoiled heavy nuclei and SNR gas dynamics, and the resulting line profile can become strongly broadened or even approach a quasi-continuum-like shape.
Figure~\ref{fig: all_spe} shows the differential de-excitation line spectra of the three target SNRs for different proton spectral indices.
In general, the total line flux increases with increasing spectral index, reflecting the enhanced contribution of low-energy protons in steeper spectra.
Distinct emission lines are clearly identified at 1.63 MeV mainly from $^{20}$Ne, 4.44 MeV mainly from $^{12}$C, and 6.13 MeV mainly from $^{16}$O.
The $6.13$ MeV line from $^{16}$O is particularly noteworthy for the SNRs considered here.
All three objects are associated with core-collapse supernovae, for which oxygen is a characteristic nucleosynthetic product of massive-star progenitors.
The appreciable O abundances inferred for our targets thus make the $^{16}$O de-excitation line a particularly relevant spectral feature for this sample, although its predicted flux also depends on the accelerated-particle spectrum and composition, the target material, and the corresponding nuclear reaction cross sections.
For the representative case of $\alpha=4.0$, including broadening due to shock motion and nuclear recoil, the  intrinsic full width at half maximum (FWHM) of the 4.44 MeV and 6.13 MeV lines in Vela are found to be approximately $\sim0.10$ MeV and $\sim0.12$ MeV, respectively.
For each feature, we define a central energy interval containing 99\% of the photons in its narrow line component from direct process, excluding 0.5\% from each tail. 
We then integrate the intrinsic model spectrum over this interval.
These integrated fluxes include contributions from overlapping spectral features and do not represent the total flux obtained by summing contributions from direct and inverse processes associated with a single nuclear transition.
To assess instrumental energy smearing, we have checked Gaussian convolution over 0.2–5 MeV, adopting an approximate COSI energy resolution of $\Delta E_{\rm FWHM}=9\,{\rm keV}\sqrt{E_\gamma/(1.157\,{\rm MeV})}$, anchored to the published COSI resolution requirements \citep{COSI2023}.
Owing to COSI’s high energy resolution, this convolution produces only minor changes to the overall line profiles.
Figures~\ref{fig: all_spe} -- \ref{fig: IC} therefore show the intrinsic nuclear emission spectra before instrumental convolution.
Finally, the last panel of Figure~\ref{fig: all_spe} shows the total line emission from the Vela SNR region. 
For this comparison, we adopt $\alpha=3.0$ for Vela, $\alpha=1.8$ Vela Jr. and  $\alpha = 2.3$ for Puppis A (see the right panel in Figure~\ref{fig: sed_vela}), consistent with the GeV -- TeV spectral studies. 
The corresponding total fluxes are $\sim1.6\times10^{-7}$, $\sim2.1\times10^{-7}$ and $\sim8.3\times10^{-8}$ $\rm{ph} \cm^{-2}\s^{-1}$ for the 1.63, 4.44 and 6.13 MeV lines, respectively.
At the same adopted proton energy budget, the BPL model yields integrated fluxes of $2.6\times10^{-8}$, $5.7\times10^{-8}$, and $2.3\times10^{-8}$ $\rm{ph} \cm^{-2}\s^{-1}$ around 1.63, 4.44, and 6.13 MeV, respectively. These are lower than the corresponding $\alpha=3.0$ values by factors of approximately 6.3, 3.5, and 3.5, illustrating the sensitivity of the predictions to the assumed low-energy spectrum.
Replacing the Vela component with the BPL model gives total fluxes of approximately $2.8\times10^{-8}$, $6.1\times10^{-8}$, and $2.5\times10^{-8}$ $\rm{ph} \cm^{-2}\s^{-1}$ around 1.63, 4.44, and 6.13 MeV, respectively.

\begin{table}
    \centering
    \caption{Integrated line fluxes of 1.63 MeV $^{20}$Ne, 4.44 MeV $^{12}$C and 6.13 MeV $^{16}$O lines.}
    \label{tab: flux}
    \begin{tabular}{cccc}
    \hline
    \hline
        Targets &   Flux (${\rm ph}\cm^{-2} \s^{-1}$)         &   Flux ($\rm {ph}\cm^{-2} \s^{-1}$)     &   Flux ($\rm {ph}\cm^{-2} \s^{-1}$) \\
                &   1.63 MeV        &   4.44 MeV        &   6.13 MeV \\
    \hline
        Vela    & $1.6\times10^{-7}$ -- $1.2\times10^{-6}$ &   $2.0\times10^{-7}$ -- $5.6\times10^{-7}$       &   $8.1\times10^{-8}$ -- $2.6\times10^{-7}$ \\
        Vela Jr.&  $2.9\times10^{-10}$ -- $3.7\times10^{-7}$   &  $9.6\times10^{-10}$ -- $1.7\times10^{-7}$      &   $4.7\times10^{-10}$ -- $1.1\times10^{-7}$ \\
        Puppis A&  $1.3\times10^{-10}$ -- $6.5\times10^{-8}$ & $3.2\times10^{-10}$ -- $2.2\times10^{-8}$      &   $1.6\times10^{-10}$ -- $1.4\times10^{-8}$ \\
        COSI reference$^a$ &  $\sim3.0\times10^{-6}$   &   $\sim8.8\times10^{-6}$    &    $\sim3.2\times10^{-5}$ \\
    \hline
    \end{tabular}
    \begin{tablenotes}
        \small
        \item $^a$ The COSI narrow-line sensitivity estimates are interpolated from Fig.~2 of \citet{COSI2023}; the value at 6.13~MeV is included for reference only, as this energy lies outside COSI's nominal 0.2--5~MeV band.
    \end{tablenotes}
\end{table}

\section{Discussion}\label{sec: discussion}

As shown in the last panel of Figure~\ref{fig: all_spe}, the emission is overwhelmingly dominated by the Vela SNR for the parameter set adopted above.
Therefore, Vela is taken as a representative case for further discussion of the MeV line emission in this region.
The continuum MeV $\gamma$-ray background consists primarily of two components: the large-scale diffuse Galactic emission and the localized continuum emission associated with individual astrophysical sources.

\subsection{Diffuse Galactic emission}

Using observations from the SPI spectrometer aboard INTEGRAL in the 0.5--8.0 MeV energy range, \citet{GCMeV_bkg} analyzed the diffuse Galactic Center (GC) emission with energy-dependent spatial templates based on the {\textbf GALPROP} model \citep{GALPROP}, covering a region of $\Delta l \times \Delta b = 95^\circ \times 95^\circ$ around the GC.
Their analysis indicates the presence of a broad continuum component in the inner Galaxy, which is consistent with inverse Compton (IC) emission from CR electrons, together with a subdominant ($\sim10\%$) bremsstrahlung contribution.
To estimate the diffuse continuum background, we adopt the power-law fit reported by \citet{GCMeV_bkg}, $I(E)=C_0(E/1\,\mathrm{MeV})^\alpha$, with $C_0=(3.1\pm0.3)\times10^{-6}\ \mathrm{ph\,cm^{-2}\,s^{-1}\,keV^{-1}\,sr^{-1}}$, and $\alpha=-1.39\pm0.09$.
However, the Vela SNR region is located well outside the Galactic Center region (l = 263.9$^\circ$, b = $-3.3^\circ$). 
As a result, the diffuse MeV background in the Vela direction is expected to be substantially lower than that in the GC, typically by a factor of about $10^{-2}$--$10^{-1}$ according to previous studies \citep[e.g., ][]{vela_bkg, GAL_bkg}.
We apply this order-of-magnitude scaling to the power-law intensity and integrate it over a circular region of $4^\circ$ radius to estimate the diffuse background toward Vela. Figure 4 shows the resulting spectrum and its $1\sigma$ uncertainty band.
The contrast between the predicted line features and the estimated diffuse continuum depends strongly on the assumed proton spectral index.
In particular, models with steeper spectra yield higher line fluxes.
Figure~4 shows the three-year survey continuum sensitivity of newASTROGAM for extragalactic point sources for reference \citep{newASTROGAM2025}.
For reference, the predicted 1.63 and 4.44 MeV line fluxes are compared with the published COSI $3\sigma$ narrow-line point-source sensitivity for a two-year survey, interpolated at the corresponding line energies and listed in Table III; both lie below these benchmarks.

\begin{figure}
    \includegraphics[width=0.8\textwidth]{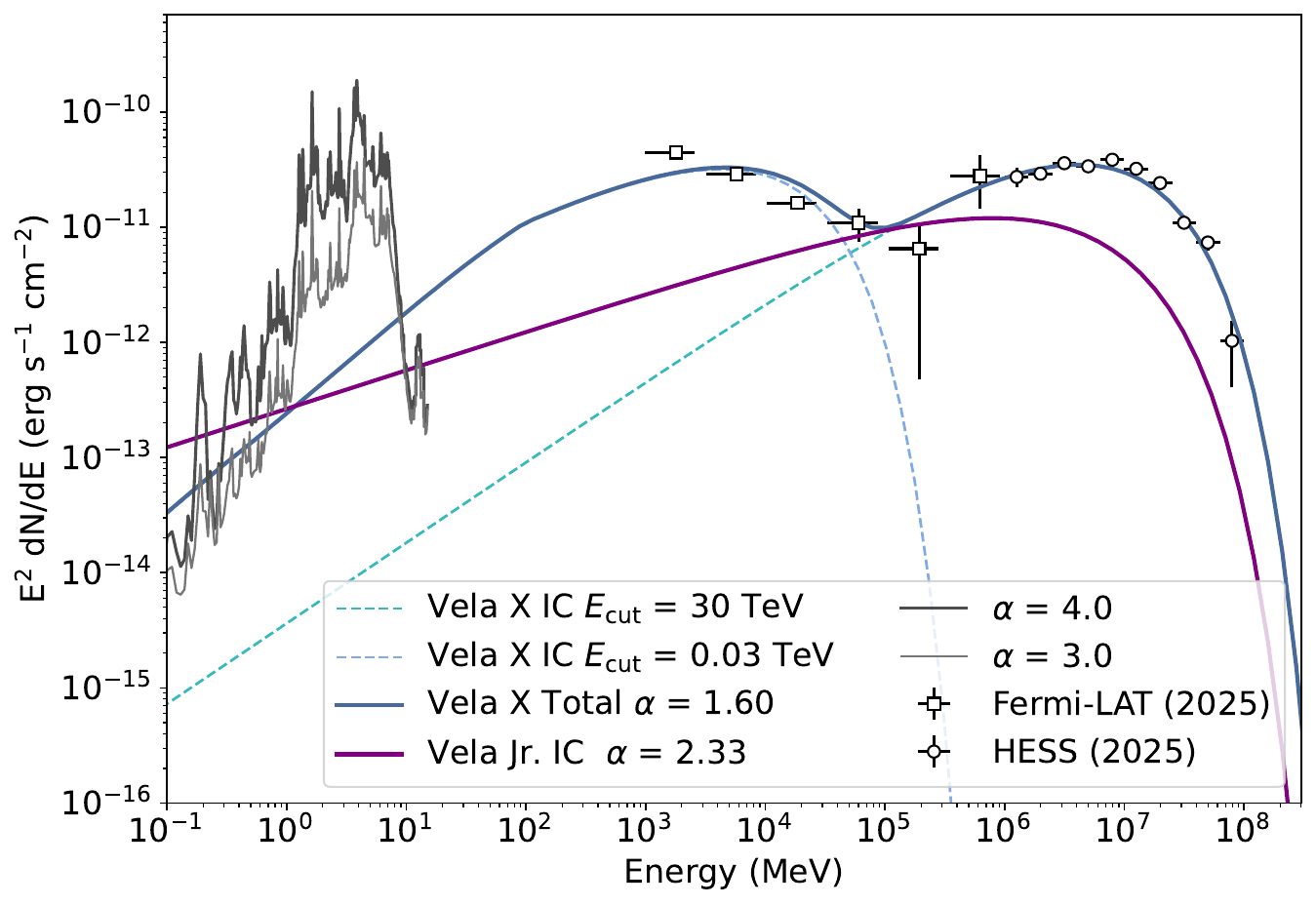}
    \caption{Predicted IC emission and MeV de-excitation line spectra in the Vela SNR region.
    The Fermi-LAT data (square points) are taken from \citet{VelaX_fermi_2025}, while the H.E.S.S. data (circle points) are adopted from the latest results presented at ICRC 2025 (see details in Section~\ref{sec: discussion}).
    For Vela X, the IC component with a low cutoff energy of $E_{\rm cut}=0.03$ TeV is shown by the light-blue dashed curve, while the model with a high cutoff energy of $E_{\rm cut}=30$ TeV is shown by the cyan dashed curve.
    The corresponding total IC emission is shown by the dark-blue curve.
    The IC emission model of Vela Jr. is shown by the purple curve and is adopted from \citet{VelaJr_gray}.
    The black and gray curves reproduce the intrinsic Vela spectra for $\alpha=4.0$ and $3.0$, respectively, shown in Figure~\ref{fig: bkg}.
    }
    \label{fig: IC}
\end{figure}

\subsection{Localized continuum emission}

Besides the diffuse Galactic background, continuum emission associated with the SNRs themselves should also be considered. 
According to the GeV -- TeV studies of \citet{puppisa_GeV}, the \gray\ emission of Puppis A is predominantly hadronic in origin. 
Therefore, its continuum contribution in the MeV band is expected to remain below, or at most comparable to, the diffuse background level.
Figure~\ref{fig: vela_radio} shows that the PWN Vela X is located inside the shell of the Vela SNR. 
The IC emission from Vela X may therefore constitute an additional background component for the detection of nuclear de-excitation lines. 
Previous \gray\ studies of the Vela X region have identified a large extended source with a radius of approximately $1.36^\circ$ \citep{VelaX_fermi_2018, VelaX_fermi_2025}. 
The spectral energy distribution exhibits evidence for multiple components.
While the spectrum is relatively soft at lower energies, a spectral hardening is observed at the highest energies covered by Fermi-LAT, providing a natural connection to the TeV emission detected by H.E.S.S.
To estimate the IC background in the MeV range, we combine the Fermi-LAT measurements \citep{VelaX_fermi_2025} with the latest H.E.S.S. results presented at ICRC 2025 \footnote{\url{https://indico.cern.ch/event/1258933/contributions/6491176/}}.
For the H.E.S.S. data, only the total flux is adopted.
Following the modeling approach of \citet{VelaX_fermi_2018}, we assume two electron populations with a common power-law index of 1.6 and cutoff energies of $E_{\rm cut}=0.03$ TeV and $E_{\rm cut}=30$ TeV, respectively.
The corresponding IC emission is then calculated, and the fitting results are shown in Figure~\ref{fig: IC}.
In addition, the GeV--TeV studies of Vela Jr. have also explored a leptonic scenario and suggested possible PWN-related emission associated with PSR J0855$-$4644 within the SNR.
The corresponding leptonic model is shown by the purple curve in Figure~\ref{fig: IC}.
For the $\alpha=3.0$ and 4.0 models, the prominent nuclear line features exceed the estimated local IC continuum.
This comparison alone does not establish detectability, which also depends on the diffuse and instrumental backgrounds and the instrument response.

In general, the predicted MeV nuclear de-excitation line emission in the Vela SNR region is highly model-dependent, as it depends on the assumed CR spectral shape, elemental abundances, and spatial distribution of particle acceleration.
Future observations with next-generation MeV telescopes will therefore be crucial for constraining the LECR population and identifying the dominant acceleration sites.
The Vela SNR region is particularly complex, covering a large angular scale of approximately $4^\circ$ in radius and containing multiple potential particle acceleration sites, including the Vela SNR, in which PWN Vela X is harbored.
Vela Jr. and Puppis A lie at different distances and are projected near the edge of the Vela shell.
At present, the $\gamma$-ray emission from these components is not fully spatially separated, which introduces additional uncertainty in the interpretation of any potential MeV line signal.
From a detectability perspective, the expected line emission must be compared against both the diffuse Galactic background and the continuum IC emission from leptonic sources in the same region.
Therefore, the detectability of MeV lines in this region strongly depends on the underlying particle spectral index and background level.
In particular, as shown in Figure~\ref{fig: bkg}, the relatively low diffuse background toward Vela, when integrated over a $4^\circ$ radius region, provides a favorable environment for identifying the predicted MeV line emission.
This capability will be essential for disentangling emission from different particle populations and for clarifying the dominant acceleration processes in this region.

\section{Summary}\label{sec: summary}

We have estimated the possible MeV nuclear de-excitation line emission produced by interactions of LECRs with the gas inside three core-collapse SNRs in the Vela SNR region under simplified assumptions. 
We find that the predicted line emission is generally dominated by the Vela SNR, particularly for soft injected proton spectra.
The detectability of these MeV line features is primarily limited by two background components: the diffuse Galactic MeV emission and continuum emission associated with high-energy sources in the region.
In particular, IC emission from the Vela X PWN may provide an additional continuum contribution that affects the observational prospects.

Under the simplified framework adopted in this work, the predicted integrated fluxes are about $1.3\times10^{-10}$ -- $1.2\times10^{-6}$, $3.2\times10^{-10}$ -- $5.6\times10^{-7}$  and $1.6\times10^{-10}$--$2.6\times10^{-7}$  $\rm{ph} \cm^{-2}\s^{-1}$ for 1.63, 4.44 and 6.13 MeV lines, respectively.
We find that the Vela SNR region represents one of the few nearby SNR complexes where MeV nuclear de-excitation lines may become competitive with continuum backgrounds for next-generation MeV instruments.
Although the predicted 1.63 and 4.44~MeV fluxes lie below the published COSI narrow-line sensitivity benchmarks, future MeV $\gamma$-ray missions with improved line sensitivity, such as newASTROGAM, may offer better prospects for detecting these features in the Vela SNR region \citep{newASTROGAM2025}.
Such observations would provide important constraints on particle acceleration and LECR populations in this complex environment, particularly in the Vela SNR.

\section*{Acknowledgments}
The authors thank Han-Xiao Chen and Xiao Zhang for helpful comments.
This work is supported by the NSFC under grants 12393852, 12121003,  and 12573047 and the Basic Research Program of Jiangsu NO.BK20252108.

\section*{Data Availability}
The nuclear reaction code {\textsc{TALYS}} (version 2.2) used in this work is publicly available at \url{https://nds.iaea.org/talys/}.
For several specific de-excitation lines, the production cross sections compiled by \citet{Murphy2009} were adopted in place of the corresponding {\textsc{TALYS}} predictions.
The Evaluated Nuclear Structure Data File (ENSDF) employed for line identification is publicly available at \url{https://www.nndc.bnl.gov/ensdfarchivals/}.
\end{CJK*}
\bibliography{MeVela}

@ARTICLE{Indriolo2015,
       author = {{Indriolo}, Nick and {Neufeld}, D.~A. and {Gerin}, M. and {Schilke}, P. and
         {Benz}, A.~O. and {Winkel}, B. and {Menten}, K.~M. and
         {Chambers}, E.~T. and {Black}, John H. and {Bruderer}, S. and
         {Falgarone}, E. and {Godard}, B. and {Goicoechea}, J.~R. and
         {Gupta}, H. and {Lis}, D.~C. and {Ossenkopf}, V. and {Persson}, C.~M. and
         {Sonnentrucker}, P. and {van der Tak}, F.~F.~S. and
         {van Dishoeck}, E.~F. and {Wolfire}, Mark G. and {Wyrowski}, F.},
        title = "{Herschel Survey of Galactic OH$^{+}$, H$_{2}$O$^{+}$, and H$_{3}$O$^{+}$: Probing the Molecular Hydrogen Fraction and Cosmic-Ray Ionization Rate}",
      journal = {ApJ},
         year = 2015,
        month = feb,
       volume = {800},
       number = {1},
          eid = {40},
        pages = {40},
          doi = {10.1088/0004-637X/800/1/40},
archivePrefix = {arXiv},
       eprint = {1412.1106},
 primaryClass = {astro-ph.GA},
       adsurl = {https://ui.adsabs.harvard.edu/abs/2015ApJ...800...40I}
}

@ARTICLE{cmz_nline,
       author = {{Liu}, Bing and {Yang}, Ruizhi},
        title = "{Constraining the low-energy cosmic ray flux in the center molecular zone from MeV nuclear deexcitation line observations}",
      journal = {\prd},
         year = 2024,
        month = sep,
       volume = {110},
       number = {6},
          eid = {063033},
        pages = {063033},
          doi = {10.1103/PhysRevD.110.063033},
archivePrefix = {arXiv},
       eprint = {2409.19350},
 primaryClass = {astro-ph.HE},
       adsurl = {https://ui.adsabs.harvard.edu/abs/2024PhRvD.110f3033L}
}

@ARTICLE{ele_abundance_VelaJr,
       author = {{Camilloni}, Francesco and {Becker}, Werner and {Predehl}, Peter and {Dennerl}, Konrad and {Freyberg}, Michael and {Mayer}, Martin G.~F. and {Sasaki}, Manami},
        title = "{SRG/eROSITA and XMM-Newton observations of Vela Jr}",
      journal = {\aap},
         year = 2023,
        month = may,
       volume = {673},
          eid = {A45},
        pages = {A45},
          doi = {10.1051/0004-6361/202245475},
archivePrefix = {arXiv},
       eprint = {2303.12686},
 primaryClass = {astro-ph.HE},
       adsurl = {https://ui.adsabs.harvard.edu/abs/2023A&A...673A..45C}
}

@ARTICLE{ele_abundance_Vela,
       author = {{Mayer}, Martin G.~F. and {Becker}, Werner and {Predehl}, Peter and {Sasaki}, Manami},
        title = "{A detailed look at the thermal and nonthermal X-ray emission from the Vela supernova remnant with SRG/eROSITA}",
      journal = {\aap},
         year = 2023,
        month = aug,
       volume = {676},
          eid = {A68},
        pages = {A68},
          doi = {10.1051/0004-6361/202346691},
archivePrefix = {arXiv},
       eprint = {2306.10975},
 primaryClass = {astro-ph.HE},
       adsurl = {https://ui.adsabs.harvard.edu/abs/2023A&A...676A..68M}
}

@ARTICLE{de_excitation_line_liub,
       author = {{Liu}, Bing and {Yang}, Rui-zhi and {Aharonian}, Felix},
        title = "{Nuclear de-excitation lines as a probe of low-energy cosmic rays}",
      journal = {\aap},
         year = 2021,
        month = feb,
       volume = {646},
          eid = {A149},
        pages = {A149},
          doi = {10.1051/0004-6361/202039977},
archivePrefix = {arXiv},
       eprint = {2101.03695},
 primaryClass = {astro-ph.HE},
       adsurl = {https://ui.adsabs.harvard.edu/abs/2021A&A...646A.149L}
}

@ARTICLE{de_excitation_casa,
       author = {{Liu}, Bing and {Yang}, Rui-zhi and {He}, Xin-yu and {Aharonian}, Felix},
        title = "{New estimation of the nuclear de-excitation line emission from the supernova remnant Cassiopeia A}",
      journal = {\mnras},
         year = 2023,
        month = oct,
       volume = {524},
       number = {4},
        pages = {5248-5253},
          doi = {10.1093/mnras/stad2165},
archivePrefix = {arXiv},
       eprint = {2307.08967},
 primaryClass = {astro-ph.HE},
       adsurl = {https://ui.adsabs.harvard.edu/abs/2023MNRAS.524.5248L}
}

@ARTICLE{RKL1979,
       author = {{Ramaty}, R. and {Kozlovsky}, B. and {Lingenfelter}, R.~E.},
        title = "{Nuclear gamma-rays from energetic particle interactions.}",
      journal = {\apjs},
         year = 1979,
        month = jul,
       volume = {40},
        pages = {487-526},
          doi = {10.1086/190596},
       adsurl = {https://ui.adsabs.harvard.edu/abs/1979ApJS...40..487R}
}

@ARTICLE{Murphy2009,
       author = {{Murphy}, R.~J. and {Kozlovsky}, B. and {Kiener}, J. and {Share}, G.~H.},
        title = "{Nuclear Gamma-Ray De-Excitation Lines and Continuum from Accelerated-Particle Interactions in Solar Flares}",
      journal = {\apjs},
         year = 2009,
        month = jul,
       volume = {183},
       number = {1},
        pages = {142-155},
          doi = {10.1088/0067-0049/183/1/142},
       adsurl = {https://ui.adsabs.harvard.edu/abs/2009ApJS..183..142M}
}

@ARTICLE{Vela_GeV,
       author = {{Araya}, Miguel and {Ram{\'\i}rez}, Santiago and {Bueso}, Diego and {Solano-Rojas}, Braulio J.},
        title = "{GeV emission in the region of Vela: a new view of the supernova remnant}",
      journal = {arXiv e-prints},
         year = 2026,
        month = apr,
          eid = {arXiv:2604.16768},
        pages = {arXiv:2604.16768},
archivePrefix = {arXiv},
       eprint = {2604.16768},
 primaryClass = {astro-ph.HE},
       adsurl = {https://ui.adsabs.harvard.edu/abs/2026arXiv260416768A}
}

@ARTICLE{abun_puppisa,
       author = {{Mayer}, Martin G.~F. and {Becker}, Werner and {Predehl}, Peter and {Sasaki}, Manami and {Freyberg}, Michael},
        title = "{A global view of shocked plasma in the supernova remnant Puppis A provided by SRG/eROSITA}",
      journal = {\aap},
         year = 2022,
        month = may,
       volume = {661},
          eid = {A31},
        pages = {A31},
          doi = {10.1051/0004-6361/202142517},
archivePrefix = {arXiv},
       eprint = {2110.12220},
 primaryClass = {astro-ph.HE},
       adsurl = {https://ui.adsabs.harvard.edu/abs/2022A&A...661A..31M}
}

@ARTICLE{puppisa_GeV,
       author = {{Giuffrida}, R. and {Lemoine-Gourmard}, M. and {Miceli}, M. and {Gabici}, S. and {Sano}, H. and {Aruga}, M. and {Mayer}, M. and {Becker}, W. and {Sasaki}, M. and {Fukui}, Y.},
        title = "{Evidence for protons accelerated and escaped from the Puppis A region using Fermi-LAT observations}",
      journal = {\aap},
         year = 2025,
        month = sep,
       volume = {701},
          eid = {A206},
        pages = {A206},
          doi = {10.1051/0004-6361/202555126},
archivePrefix = {arXiv},
       eprint = {2507.11208},
 primaryClass = {astro-ph.HE},
       adsurl = {https://ui.adsabs.harvard.edu/abs/2025A&A...701A.206G}
}

@ARTICLE{VelaJr_gray,
       author = {{H.~E.~S.~S. Collaboration} and {Abdalla}, H. and {Abramowski}, A. and {Aharonian}, F. and {Ait Benkhali}, F. and {Akhperjanian}, A.~G. and {Ang{\"u}ner}, E.~O. and {Arakawa}, M. and {Arrieta}, M. and {Aubert}, P. and {Backes}, M. and {Balzer}, A. and {Barnard}, M. and {Becherini}, Y. and {Becker Tjus}, J. and {Berge}, D. and {Bernhard}, S. and {Bernl{\"o}hr}, K. and {Blackwell}, R. and {B{\"o}ttcher}, M. and {Boisson}, C. and {Bolmont}, J. and {Bordas}, P. and {Bregeon}, J. and {Brun}, F. and {Brun}, P. and {Bryan}, M. and {B{\"u}chele}, M. and {Bulik}, T. and {Capasso}, M. and {Carr}, J. and {Casanova}, S. and {Cerruti}, M. and {Chakraborty}, N. and {Chalme-Calvet}, R. and {Chaves}, R.~C.~G. and {Chen}, A. and {Chevalier}, J. and {Chr{\'e}tien}, M. and {Coffaro}, M. and {Colafrancesco}, S. and {Cologna}, G. and {Condon}, B. and {Conrad}, J. and {Cui}, Y. and {Davids}, I.~D. and {Decock}, J. and {Degrange}, B. and {Deil}, C. and {Devin}, J. and {deWilt}, P. and {Dirson}, L. and {Djannati-Ata{\"\i}}, A. and {Domainko}, W. and {Donath}, A. and {Drury}, L.~O. 'C. and {Dutson}, K. and {Dyks}, J. and {Edwards}, T. and {Egberts}, K. and {Eger}, P. and {Ernenwein}, J.-P. and {Eschbach}, S. and {Farnier}, C. and {Fegan}, S. and {Fernandes}, M.~V. and {Fiasson}, A. and {Fontaine}, G. and {F{\"o}rster}, A. and {Funk}, S. and {F{\"u}{\ss}ling}, M. and {Gabici}, S. and {Gajdus}, M. and {Gallant}, Y.~A. and {Garrigoux}, T. and {Giavitto}, G. and {Giebels}, B. and {Glicenstein}, J.~F. and {Gottschall}, D. and {Goyal}, A. and {Grondin}, M.-H. and {Hahn}, J. and {Haupt}, M. and {Hawkes}, J. and {Heinzelmann}, G. and {Henri}, G. and {Hermann}, G. and {Hervet}, O. and {Hinton}, J.~A. and {Hofmann}, W. and {Hoischen}, C. and {Holler}, M. and {Horns}, D. and {Ivascenko}, A. and {Iwasaki}, H. and {Jacholkowska}, A. and {Jamrozy}, M. and {Janiak}, M. and {Jankowsky}, D. and {Jankowsky}, F. and {Jingo}, M. and {Jogler}, T. and {Jouvin}, L. and {Jung-Richardt}, I. and {Kastendieck}, M.~A. and {Katarzy{\'n}ski}, K. and {Katsuragawa}, M. and {Katz}, U. and {Kerszberg}, D. and {Khangulyan}, D. and {Kh{\'e}lifi}, B. and {Kieffer}, M. and {King}, J. and {Klepser}, S. and {Klochkov}, D. and {Klu{\'z}niak}, W. and {Kolitzus}, D. and {Komin}, Nu. and {Krakau}, S. and {Kraus}, M. and {Kr{\"u}ger}, P.~P. and {Laffon}, H. and {Lamanna}, G. and {Lau}, J. and {Lees}, J.-P. and {Lefaucheur}, J. and {Lefranc}, V. and {Lemi{\`e}re}, A. and {Lemoine-Goumard}, M. and {Lenain}, J.-P. and {Leser}, E. and {Lohse}, T. and {Lorentz}, M. and {Liu}, R. and {L{\'o}pez-Coto}, R. and {Lypova}, I. and {Marandon}, V. and {Marcowith}, A. and {Mariaud}, C. and {Marx}, R. and {Maurin}, G. and {Maxted}, N. and {Mayer}, M. and {Meintjes}, P.~J. and {Meyer}, M. and {Mitchell}, A.~M.~W. and {Moderski}, R. and {Mohamed}, M. and {Mohrmann}, L. and {Mor{\r{a}}}, K. and {Moulin}, E. and {Murach}, T. and {Nakashima}, S. and {de Naurois}, M. and {Niederwanger}, F. and {Niemiec}, J. and {Oakes}, L. and {O'Brien}, P. and {Odaka}, H. and {{\"O}ttl}, S. and {Ohm}, S. and {Ostrowski}, M. and {Oya}, I. and {Padovani}, M. and {Panter}, M. and {Parsons}, R.~D. and {Paz Arribas}, M. and {Pekeur}, N.~W. and {Pelletier}, G. and {Perennes}, C. and {Petrucci}, P.-O. and {Peyaud}, B. and {Piel}, Q. and {Pita}, S. and {Poon}, H. and {Prokhorov}, D. and {Prokoph}, H. and {P{\"u}hlhofer}, G. and {Punch}, M. and {Quirrenbach}, A. and {Raab}, S. and {Reimer}, A. and {Reimer}, O. and {Renaud}, M. and {de los Reyes}, R. and {Richter}, S. and {Rieger}, F. and {Romoli}, C. and {Rowell}, G. and {Rudak}, B. and {Rulten}, C.~B. and {Sahakian}, V. and {Saito}, S. and {Salek}, D. and {Sanchez}, D.~A. and {Santangelo}, A. and {Sasaki}, M. and {Schlickeiser}, R. and {Sch{\"u}ssler}, F. and {Schulz}, A.},
        title = "{Deeper H.E.S.S. observations of Vela Junior (RX J0852.0-4622): Morphology studies and resolved spectroscopy}",
      journal = {\aap},
         year = 2018,
        month = apr,
       volume = {612},
          eid = {A7},
        pages = {A7},
          doi = {10.1051/0004-6361/201630002},
archivePrefix = {arXiv},
       eprint = {1611.01863},
 primaryClass = {astro-ph.HE},
       adsurl = {https://ui.adsabs.harvard.edu/abs/2018A&A...612A...7H}
}

@ARTICLE{Vela_radio,
       author = {{Duncan}, A.~R. and {Stewart}, R.~T. and {Haynes}, R.~F. and {Jones}, K.~L.},
        title = "{The VELA supernova remnant and the GUM nebula: new perspectives at 2.4 GHz}",
      journal = {\mnras},
         year = 1996,
        month = may,
       volume = {280},
       number = {1},
        pages = {252-266},
          doi = {10.1093/mnras/280.1.252},
       adsurl = {https://ui.adsabs.harvard.edu/abs/1996MNRAS.280..252D}
}

@ARTICLE{velaage1,
       author = {{van der Swaluw}, E. and {Wu}, Y.},
        title = "{Inferring Initial Spin Periods for Neutron Stars in Composite Remnants}",
      journal = {\apjl},
         year = 2001,
        month = jul,
       volume = {555},
       number = {1},
        pages = {L49-L53},
          doi = {10.1086/321733},
archivePrefix = {arXiv},
       eprint = {astro-ph/0104390},
 primaryClass = {astro-ph},
       adsurl = {https://ui.adsabs.harvard.edu/abs/2001ApJ...555L..49V}
}

@INPROCEEDINGS{velaage2,
       author = {{Kargaltsev}, O. and {Pavlov}, G.~G.},
        title = "{Pulsar Wind Nebulae in the Chandra Era}",
    booktitle = {40 Years of Pulsars: Millisecond Pulsars, Magnetars and More},
         year = 2008,
       editor = {{Bassa}, C. and {Wang}, Z. and {Cumming}, A. and {Kaspi}, V.~M.},
       series = {American Institute of Physics Conference Series},
       volume = {983},
        month = feb,
    publisher = {AIP},
        pages = {171-185},
          doi = {10.1063/1.2900138},
archivePrefix = {arXiv},
       eprint = {0801.2602},
 primaryClass = {astro-ph},
       adsurl = {https://ui.adsabs.harvard.edu/abs/2008AIPC..983..171K}
}

@ARTICLE{velaage3,
       author = {{Suzuki}, Hiromasa and {Bamba}, Aya and {Shibata}, Shinpei},
        title = "{Quantitative Age Estimation of Supernova Remnants and Associated Pulsars}",
      journal = {\apj},
         year = 2021,
        month = jun,
       volume = {914},
       number = {2},
          eid = {103},
        pages = {103},
          doi = {10.3847/1538-4357/abfb02},
archivePrefix = {arXiv},
       eprint = {2104.10052},
 primaryClass = {astro-ph.HE},
       adsurl = {https://ui.adsabs.harvard.edu/abs/2021ApJ...914..103S}
}

@ARTICLE{Velajr_age,
       author = {{Allen}, G.~E. and {Chow}, K. and {DeLaney}, T. and {Filipovi{\'c}}, M.~D. and {Houck}, J.~C. and {Pannuti}, T.~G. and {Stage}, M.~D.},
        title = "{On the Expansion Rate, Age, and Distance of the Supernova Remnant G266.2-1.2 (Vela Jr.)}",
      journal = {\apj},
         year = 2015,
        month = jan,
       volume = {798},
       number = {2},
          eid = {82},
        pages = {82},
          doi = {10.1088/0004-637X/798/2/82},
archivePrefix = {arXiv},
       eprint = {1410.7435},
 primaryClass = {astro-ph.HE},
       adsurl = {https://ui.adsabs.harvard.edu/abs/2015ApJ...798...82A}
}

@ARTICLE{PuppisA_age,
       author = {{Leahy}, D.~A. and {Ranasinghe}, S. and {Gelowitz}, M.},
        title = "{Evolutionary Models for 43 Galactic Supernova Remnants with Distances and X-Ray Spectra}",
      journal = {\apjs},
         year = 2020,
        month = may,
       volume = {248},
       number = {1},
          eid = {16},
        pages = {16},
          doi = {10.3847/1538-4365/ab8bd9},
archivePrefix = {arXiv},
       eprint = {2003.08998},
 primaryClass = {astro-ph.HE},
       adsurl = {https://ui.adsabs.harvard.edu/abs/2020ApJS..248...16L}
}

@ARTICLE{Vela_dis,
       author = {{Cha}, Alexandra N. and {Sembach}, Kenneth R. and {Danks}, Anthony C.},
        title = "{The Distance to the Vela Supernova Remnant}",
      journal = {\apjl},
         year = 1999,
        month = apr,
       volume = {515},
       number = {1},
        pages = {L25-L28},
          doi = {10.1086/311968},
archivePrefix = {arXiv},
       eprint = {astro-ph/9902230},
 primaryClass = {astro-ph},
       adsurl = {https://ui.adsabs.harvard.edu/abs/1999ApJ...515L..25C}
}

@ARTICLE{velajr_dis,
       author = {{Katsuda}, S. and {Tsunemi}, H. and {Mori}, K.},
        title = "{The Slow X-Ray Expansion of the Northwestern Rim of the Supernova Remnant RX J0852.0-4622}",
      journal = {\apjl},
         year = 2008,
        month = may,
       volume = {678},
       number = {1},
        pages = {L35},
          doi = {10.1086/588499},
archivePrefix = {arXiv},
       eprint = {0803.3266},
 primaryClass = {astro-ph},
       adsurl = {https://ui.adsabs.harvard.edu/abs/2008ApJ...678L..35K}
}

@ARTICLE{voyager,
       author = {{Cummings}, A.~C. and {Stone}, E.~C. and {Heikkila}, B.~C. and {Lal}, N. and {Webber}, W.~R. and {J{\'o}hannesson}, G. and {Moskalenko}, I.~V. and {Orlando}, E. and {Porter}, T.~A.},
        title = "{Galactic Cosmic Rays in the Local Interstellar Medium: Voyager 1 Observations and Model Results}",
      journal = {\apj},
         year = 2016,
        month = nov,
       volume = {831},
       number = {1},
          eid = {18},
        pages = {18},
          doi = {10.3847/0004-637X/831/1/18},
       adsurl = {https://ui.adsabs.harvard.edu/abs/2016ApJ...831...18C}
}

@ARTICLE{talys,
       author = {{Koning}, Arjan and {Hilaire}, Stephane and {Goriely}, Stephane},
        title = "{TALYS: modeling of nuclear reactions}",
      journal = {European Physical Journal A},
         year = 2023,
        month = jun,
       volume = {59},
       number = {6},
          eid = {131},
        pages = {131},
          doi = {10.1140/epja/s10050-023-01034-3},
       adsurl = {https://ui.adsabs.harvard.edu/abs/2023EPJA...59..131K}
}

@ARTICLE{BKT2013,
       author = {{Benhabiles-Mezhoud}, H. and {Kiener}, J. and {Tatischeff}, V. and {Strong}, A.~W.},
        title = "{De-excitation Nuclear Gamma-Ray Line Emission from Low-energy Cosmic Rays in the Inner Galaxy}",
      journal = {\apj},
         year = 2013,
        month = feb,
       volume = {763},
       number = {2},
          eid = {98},
        pages = {98},
          doi = {10.1088/0004-637X/763/2/98},
archivePrefix = {arXiv},
       eprint = {1212.1622},
 primaryClass = {astro-ph.HE},
       adsurl = {https://ui.adsabs.harvard.edu/abs/2013ApJ...763...98B}
}

@ARTICLE{GCMeV_bkg,
       author = {{Siegert}, Thomas and {Berteaud}, Joanna and {Calore}, Francesca and {Serpico}, Pasquale D. and {Weinberger}, Christoph},
        title = "{Diffuse Galactic emission spectrum between 0.5 and 8.0 MeV}",
      journal = {\aap},
         year = 2022,
        month = apr,
       volume = {660},
          eid = {A130},
        pages = {A130},
          doi = {10.1051/0004-6361/202142639},
archivePrefix = {arXiv},
       eprint = {2202.04574},
 primaryClass = {astro-ph.HE},
       adsurl = {https://ui.adsabs.harvard.edu/abs/2022A&A...660A.130S}
}

@ARTICLE{GALPROP,
       author = {{Vladimirov}, A.~E. and {Digel}, S.~W. and {J{\'o}hannesson}, G. and {Michelson}, P.~F. and {Moskalenko}, I.~V. and {Nolan}, P.~L. and {Orlando}, E. and {Porter}, T.~A. and {Strong}, A.~W.},
        title = "{GALPROP WebRun: An internet-based service for calculating galactic cosmic ray propagation and associated photon emissions}",
      journal = {Computer Physics Communications},
         year = 2011,
        month = may,
       volume = {182},
       number = {5},
        pages = {1156-1161},
          doi = {10.1016/j.cpc.2011.01.017},
archivePrefix = {arXiv},
       eprint = {1008.3642},
 primaryClass = {astro-ph.HE},
       adsurl = {https://ui.adsabs.harvard.edu/abs/2011CoPhC.182.1156V}
}

@ARTICLE{vela_bkg,
       author = {{Porter}, T.~A. and {J{\'o}hannesson}, G. and {Moskalenko}, I.~V.},
        title = "{High-energy Gamma Rays from the Milky Way: Three-dimensional Spatial Models for the Cosmic-Ray and Radiation Field Densities in the Interstellar Medium}",
      journal = {\apj},
         year = 2017,
        month = sep,
       volume = {846},
       number = {1},
          eid = {67},
        pages = {67},
          doi = {10.3847/1538-4357/aa844d},
archivePrefix = {arXiv},
       eprint = {1708.00816},
 primaryClass = {astro-ph.HE},
       adsurl = {https://ui.adsabs.harvard.edu/abs/2017ApJ...846...67P}
}

@INPROCEEDINGS{GAL_bkg,
       author = {{Strong}, A.~W. and {Moskalenko}, I.~V.},
        title = "{A 3D time-dependent model for Galactic cosmic rays and gamma rays}",
    booktitle = {International Cosmic Ray Conference},
         year = 2001,
       series = {International Cosmic Ray Conference},
       volume = {5},
        month = aug,
        pages = {1964-1967},
          doi = {10.48550/arXiv.astro-ph/0106505},
archivePrefix = {arXiv},
       eprint = {astro-ph/0106505},
 primaryClass = {astro-ph},
       adsurl = {https://ui.adsabs.harvard.edu/abs/2001ICRC....5.1964S}
}

@ARTICLE{VelaX_fermi_2018,
       author = {{Tibaldo}, L. and {Zanin}, R. and {Faggioli}, G. and {Ballet}, J. and {Grondin}, M.-H. and {Hinton}, J.~A. and {Lemoine-Goumard}, M.},
        title = "{Disentangling multiple high-energy emission components in the Vela X pulsar wind nebula with the Fermi Large Area Telescope}",
      journal = {\aap},
         year = 2018,
        month = sep,
       volume = {617},
          eid = {A78},
        pages = {A78},
          doi = {10.1051/0004-6361/201833356},
archivePrefix = {arXiv},
       eprint = {1806.11499},
 primaryClass = {astro-ph.HE},
       adsurl = {https://ui.adsabs.harvard.edu/abs/2018A&A...617A..78T}
}

@ARTICLE{VelaX_fermi_2025,
       author = {{Lange}, Alexander and {Eagle}, J. and {Kargaltsev}, O. and {Kuiper}, Lucien and {Hare}, Jeremy},
        title = "{The Vela Pulsar and Its Pulsar Wind Nebula Vela-X Using 13 Yr of Fermi-LAT Observations}",
      journal = {\apj},
         year = 2025,
        month = aug,
       volume = {988},
       number = {2},
          eid = {200},
        pages = {200},
          doi = {10.3847/1538-4357/ade713},
archivePrefix = {arXiv},
       eprint = {2506.16687},
 primaryClass = {astro-ph.HE},
       adsurl = {https://ui.adsabs.harvard.edu/abs/2025ApJ...988..200L}
}

@ARTICLE{CR280MeV,
       author = {{Schuppan}, F. and {R{\"o}ken}, C. and {Fedrau}, N. and {Becker Tjus}, J.},
        title = "{Ionisation as indicator for cosmic ray acceleration}",
      journal = {ASTRA Proceedings},
         year = 2014,
        month = jun,
       volume = {1},
       number = {1},
        pages = {13-17},
          doi = {10.5194/ap-1-13-2014},
archivePrefix = {arXiv},
       eprint = {1406.0974},
 primaryClass = {astro-ph.HE},
       adsurl = {https://ui.adsabs.harvard.edu/abs/2014ASTRP...1...13S}
}

@INPROCEEDINGS{ic443,
       author = {{Indriolo}, Nick and {Fields}, Brian D. and {McCall}, Benjamin J.},
        title = "{Probing cosmic-ray acceleration and propagation with H$_{3}$$^{+}$ observations}",
    booktitle = {Computational Methods in Sciences and Engineering 2010 (ICCMSE-2010)},
         year = 2015,
       series = {American Institute of Physics Conference Series},
       volume = {1642},
        month = jan,
    publisher = {AIP},
        pages = {370-372},
          doi = {10.1063/1.4906696},
       adsurl = {https://ui.adsabs.harvard.edu/abs/2015AIPC.1642..370I}
}

@ARTICLE{W28_V,
       author = {{Vaupr{\'e}}, S. and {Hily-Blant}, P. and {Ceccarelli}, C. and {Dubus}, G. and {Gabici}, S. and {Montmerle}, T.},
        title = "{Cosmic ray induced ionisation of a molecular cloud shocked by the W28 supernova remnant}",
      journal = {\aap},
         year = 2014,
        month = aug,
       volume = {568},
          eid = {A50},
        pages = {A50},
          doi = {10.1051/0004-6361/201424036},
archivePrefix = {arXiv},
       eprint = {1407.0205},
 primaryClass = {astro-ph.GA},
       adsurl = {https://ui.adsabs.harvard.edu/abs/2014A&A...568A..50V}
}

@ARTICLE{W28_Tu,
       author = {{Tu}, Tian-yu and {Chen}, Yang and {Zhou}, Ping and {Safi-Harb}, Samar and {Liu}, Qian-Cheng},
        title = "{Shock and Cosmic-Ray Chemistry Associated with the Supernova Remnant W28}",
      journal = {\apj},
         year = 2024,
        month = may,
       volume = {966},
       number = {2},
          eid = {178},
        pages = {178},
          doi = {10.3847/1538-4357/ad3634},
archivePrefix = {arXiv},
       eprint = {2403.13305},
 primaryClass = {astro-ph.GA},
       adsurl = {https://ui.adsabs.harvard.edu/abs/2024ApJ...966..178T}
}

@PHDTHESIS{280MeV_chem,
       author = {{Vaupre}, Solenn},
        title = "{Cosmic-ray ionisation of dense molecular clouds}",
       school = {Universite Joseph Fourier Grenoble, France},
         year = 2015,
        month = jul,
       adsurl = {https://ui.adsabs.harvard.edu/abs/2015PhDT........24V}
}

@ARTICLE{w49b,
       author = {{Zhou}, Ping and {Zhang}, Gao-Yuan and {Zhou}, Xin and {Arias}, Maria and {Koo}, Bon-Chul and {Vink}, Jacco and {Zhang}, Zhi-Yu and {Sun}, Lei and {Du}, Fu-Jun and {Zhu}, Hui and {Chen}, Yang and {Bovino}, Stefano and {Lee}, Yong-Hyun},
        title = "{Unusually High HCO$^{+}$/CO Ratios in and outside Supernova Remnant W49B}",
      journal = {\apj},
         year = 2022,
        month = jun,
       volume = {931},
       number = {2},
          eid = {144},
        pages = {144},
          doi = {10.3847/1538-4357/ac63b5},
archivePrefix = {arXiv},
       eprint = {2203.13111},
 primaryClass = {astro-ph.GA},
       adsurl = {https://ui.adsabs.harvard.edu/abs/2022ApJ...931..144Z}
}

@ARTICLE{IC443_W44_gray,
       author = {{Ackermann}, M. and {Ajello}, M. and {Allafort}, A. and {Baldini}, L. and {Ballet}, J. and {Barbiellini}, G. and {Baring}, M.~G. and {Bastieri}, D. and {Bechtol}, K. and {Bellazzini}, R. and {Blandford}, R.~D. and {Bloom}, E.~D. and {Bonamente}, E. and {Borgland}, A.~W. and {Bottacini}, E. and {Brandt}, T.~J. and {Bregeon}, J. and {Brigida}, M. and {Bruel}, P. and {Buehler}, R. and {Busetto}, G. and {Buson}, S. and {Caliandro}, G.~A. and {Cameron}, R.~A. and {Caraveo}, P.~A. and {Casandjian}, J.~M. and {Cecchi}, C. and {{\c{C}}elik}, {\"O}. and {Charles}, E. and {Chaty}, S. and {Chaves}, R.~C.~G. and {Chekhtman}, A. and {Cheung}, C.~C. and {Chiang}, J. and {Chiaro}, G. and {Cillis}, A.~N. and {Ciprini}, S. and {Claus}, R. and {Cohen-Tanugi}, J. and {Cominsky}, L.~R. and {Conrad}, J. and {Corbel}, S. and {Cutini}, S. and {D'Ammando}, F. and {de Angelis}, A. and {de Palma}, F. and {Dermer}, C.~D. and {do Couto e Silva}, E. and {Drell}, P.~S. and {Drlica-Wagner}, A. and {Falletti}, L. and {Favuzzi}, C. and {Ferrara}, E.~C. and {Franckowiak}, A. and {Fukazawa}, Y. and {Funk}, S. and {Fusco}, P. and {Gargano}, F. and {Germani}, S. and {Giglietto}, N. and {Giommi}, P. and {Giordano}, F. and {Giroletti}, M. and {Glanzman}, T. and {Godfrey}, G. and {Grenier}, I.~A. and {Grondin}, M.-H. and {Grove}, J.~E. and {Guiriec}, S. and {Hadasch}, D. and {Hanabata}, Y. and {Harding}, A.~K. and {Hayashida}, M. and {Hayashi}, K. and {Hays}, E. and {Hewitt}, J.~W. and {Hill}, A.~B. and {Hughes}, R.~E. and {Jackson}, M.~S. and {Jogler}, T. and {J{\'o}hannesson}, G. and {Johnson}, A.~S. and {Kamae}, T. and {Kataoka}, J. and {Katsuta}, J. and {Kn{\"o}dlseder}, J. and {Kuss}, M. and {Lande}, J. and {Larsson}, S. and {Latronico}, L. and {Lemoine-Goumard}, M. and {Longo}, F. and {Loparco}, F. and {Lovellette}, M.~N. and {Lubrano}, P. and {Madejski}, G.~M. and {Massaro}, F. and {Mayer}, M. and {Mazziotta}, M.~N. and {McEnery}, J.~E. and {Mehault}, J. and {Michelson}, P.~F. and {Mignani}, R.~P. and {Mitthumsiri}, W. and {Mizuno}, T. and {Moiseev}, A.~A. and {Monzani}, M.~E. and {Morselli}, A. and {Moskalenko}, I.~V. and {Murgia}, S. and {Nakamori}, T. and {Nemmen}, R. and {Nuss}, E. and {Ohno}, M. and {Ohsugi}, T. and {Omodei}, N. and {Orienti}, M. and {Orlando}, E. and {Ormes}, J.~F. and {Paneque}, D. and {Perkins}, J.~S. and {Pesce-Rollins}, M. and {Piron}, F. and {Pivato}, G. and {Rain{\`o}}, S. and {Rando}, R. and {Razzano}, M. and {Razzaque}, S. and {Reimer}, A. and {Reimer}, O. and {Ritz}, S. and {Romoli}, C. and {S{\'a}nchez-Conde}, M. and {Schulz}, A. and {Sgr{\`o}}, C. and {Simeon}, P.~E. and {Siskind}, E.~J. and {Smith}, D.~A. and {Spandre}, G. and {Spinelli}, P. and {Stecker}, F.~W. and {Strong}, A.~W. and {Suson}, D.~J. and {Tajima}, H. and {Takahashi}, H. and {Takahashi}, T. and {Tanaka}, T. and {Thayer}, J.~G. and {Thayer}, J.~B. and {Thompson}, D.~J. and {Thorsett}, S.~E. and {Tibaldo}, L. and {Tibolla}, O. and {Tinivella}, M. and {Troja}, E. and {Uchiyama}, Y. and {Usher}, T.~L. and {Vandenbroucke}, J. and {Vasileiou}, V. and {Vianello}, G. and {Vitale}, V. and {Waite}, A.~P. and {Werner}, M. and {Winer}, B.~L. and {Wood}, K.~S. and {Wood}, M. and {Yamazaki}, R. and {Yang}, Z. and {Zimmer}, S.},
        title = "{Detection of the Characteristic Pion-Decay Signature in Supernova Remnants}",
      journal = {Science},
         year = 2013,
        month = feb,
       volume = {339},
       number = {6121},
        pages = {807-811},
          doi = {10.1126/science.1231160},
archivePrefix = {arXiv},
       eprint = {1302.3307},
 primaryClass = {astro-ph.HE},
       adsurl = {https://ui.adsabs.harvard.edu/abs/2013Sci...339..807A}
}

@ARTICLE{loss_rate,
       author = {{Padovani}, M. and {Galli}, D. and {Glassgold}, A.~E.},
        title = "{Cosmic-ray ionization of molecular clouds}",
      journal = {\aap},
         year = 2009,
        month = jul,
       volume = {501},
       number = {2},
        pages = {619-631},
          doi = {10.1051/0004-6361/200911794},
archivePrefix = {arXiv},
       eprint = {0904.4149},
 primaryClass = {astro-ph.SR},
       adsurl = {https://ui.adsabs.harvard.edu/abs/2009A&A...501..619P}
}

@ARTICLE{VelaX_Bao1,
       author = {{Bao}, Yiwei and {Liu}, Siming and {Chen}, Yang},
        title = "{On the Gamma-Ray Nebula of Vela Pulsar. I. Very Slow Diffusion of Energetic Electrons within the TeV Nebula}",
      journal = {\apj},
         year = 2019,
        month = may,
       volume = {877},
       number = {1},
          eid = {54},
        pages = {54},
          doi = {10.3847/1538-4357/ab1908},
archivePrefix = {arXiv},
       eprint = {1907.02037},
 primaryClass = {astro-ph.HE},
       adsurl = {https://ui.adsabs.harvard.edu/abs/2019ApJ...877...54B}
}

@ARTICLE{VelaX_Bao2,
       author = {{Bao}, Yiwei and {Chen}, Yang},
        title = "{On the Gamma-Ray Nebula of Vela Pulsar. II. The Soft Spectrum of the Extended Radio Nebula}",
      journal = {\apj},
         year = 2019,
        month = aug,
       volume = {881},
       number = {2},
          eid = {148},
        pages = {148},
          doi = {10.3847/1538-4357/ab2ed8},
archivePrefix = {arXiv},
       eprint = {1907.02038},
 primaryClass = {astro-ph.HE},
       adsurl = {https://ui.adsabs.harvard.edu/abs/2019ApJ...881..148B}
}

@ARTICLE{Vela_dens2,
       author = {{Miceli}, M. and {Orlando}, S. and {Reale}, F. and {Bocchino}, F. and {Peres}, G.},
        title = "{Hydrodynamic modelling of ejecta shrapnel in the Vela supernova remnant}",
      journal = {\mnras},
         year = 2013,
        month = apr,
       volume = {430},
       number = {4},
        pages = {2864-2872},
          doi = {10.1093/mnras/stt093},
archivePrefix = {arXiv},
       eprint = {1301.3085},
 primaryClass = {astro-ph.HE},
       adsurl = {https://ui.adsabs.harvard.edu/abs/2013MNRAS.430.2864M}
}

@ARTICLE{puppisa_dis,
       author = {{Reynoso}, E.~M. and {Cichowolski}, S. and {Walsh}, A.~J.},
        title = "{A high-resolution H I study towards the supernova remnant Puppis A and its environments}",
      journal = {\mnras},
         year = 2017,
        month = jan,
       volume = {464},
       number = {3},
        pages = {3029-3039},
          doi = {10.1093/mnras/stw2219},
archivePrefix = {arXiv},
       eprint = {1609.01779},
 primaryClass = {astro-ph.GA},
       adsurl = {https://ui.adsabs.harvard.edu/abs/2017MNRAS.464.3029R}
}

@INPROCEEDINGS{newASTROGAM2025,
       author = {{Berge}, D. and {Mazziotta}, M.~N. and {Tavani}, M. and {Tatischeff}, V. and {Oberlack}, U. and {Newastrogam Proposal Team}},
        title = "{newASTROGAM: The New MeV to GeV Gamma-ray Observatory}",
    booktitle = {39th International Cosmic Ray Conference},
         year = 2025,
        month = dec,
    publisher = {SISSA},
          eid = {572},
        pages = {572},
       adsurl = {https://ui.adsabs.harvard.edu/abs/2025icrc.confE.572B}
}

@INPROCEEDINGS{COSI2023,
       author = {{Tomsick}, J. and {COSI Collaboration}},
        title = "{The Compton Spectrometer and Imager Project for MeV Astronomy}",
    booktitle = {37th International Cosmic Ray Conference},
         year = 2022,
        month = mar,
    publisher = {SISSA},
          eid = {652},
        pages = {652},
          doi = {10.22323/1.395.0652},
archivePrefix = {arXiv},
       eprint = {2109.10403},
 primaryClass = {astro-ph.IM},
       adsurl = {https://ui.adsabs.harvard.edu/abs/2022icrc.confE.652T}
}

\end{document}